# CONRep: Uncertainty-Aware Vision-Language Report Drafting Using Conformal Prediction

Danial Elyassirad, MD[1], Benyamin Gheiji[1], Mahsa Vatanparast, MD[1],

Amir Mahmoud Ahmadzadeh, MD[2], Seyed Amir Asef Agah, MD[3,4], Mana Moassefi, MD[5],

Meysam Tavakoli, PhD[6], Shahriar Faghani, MD[7,8*]

(1) Student Research Committee, Faculty of Medicine, Mashhad University of Medical Sciences, Mashhad, Iran

(2) Department of Radiology, School of Medicine, Mashhad University of Medical Sciences, Mashhad, Iran

(3) School of Medicine, Tehran University of Medical Sciences, Tehran, Iran

(4) Advanced Diagnostic and Interventional Radiology Research Center, Tehran University of Medical Sciences, Tehran, Iran

(5) Department of Radiology, Mayo Clinic, Rochester, MN, USA

(6) Department of Radiation Oncology and Winship Cancer Institute, Emory University, Atlanta, Georgia, USA

(7) Radiology Informatics Lab, Department of Radiology, Mayo Clinic, Rochester, MN, USA

(8) Department of Radiology, University of Pennsylvania, Philadelphia, PA, USA

(*) Correspondence: Shahriar Faghani, Email: Faghani.Shahriar@mayo.edu

## *Abstract*

Automated radiology report drafting (ARRD) using vision-language models (VLMs) has advanced rapidly, yet most systems lack explicit uncertainty estimates, limiting trust and safe clinical deployment. We propose CONRep, a model-agnostic framework that integrates conformal prediction (CP) to provide statistically grounded uncertainty quantification for VLM-generated radiology reports. CONRep operates at both the label level, by calibrating binary predictions for predefined findings, and the sentence level, by assessing uncertainty in free-text impressions via image-text semantic alignment. We evaluate CONRep using both generative and contrastive VLMs on public chest X-ray datasets. Across both settings, outputs classified as high confidence consistently show significantly higher agreement with radiologist annotations and ground-truth impressions than low-confidence outputs. By enabling calibrated confidence stratification without modifying underlying models, CONRep improves the transparency, reliability, and clinical usability of automated radiology reporting systems.

**Introduction**

Automated radiology report drafting (ARRD) is an emerging application of vision language models (VLMs) that aim to assist radiologists by automatically drafting structured and narrative diagnostic reports. ARRD systems have the potential to reduce reporting time, alleviate radiologist workload, and improve report consistency across institutions and practitioners [1]. Consequently, ARRD has attracted increasing interest as a promising component of future clinical reporting workflows.

Recent progress in ARRD has been driven by advances in VLMs that jointly learn representations of images and text. By aligning visual features with linguistic concepts in a shared embedding space, VLM-based approaches enable end-to-end report drafting with improved semantic grounding and expressiveness [2-4]. VLM pretraining strategies have shown substantial benefits for both image understanding and text generation in the medical domain, supporting more coherent and clinically relevant reports. Building on these foundations, large-scale biomedical VLMs have further enhanced generalization capabilities, enabling zero-shot and few-shot report generation across diverse imaging scenarios [5-11]. These developments have substantially improved the quality, flexibility, and scalability of ARRD systems.

Despite these advances, most ARRD systems produce outputs without explicit estimates of prediction confidence. In clinical practice, however, radiological interpretation is inherently uncertain due to ambiguous imaging findings, overlapping disease patterns, and inter-observer variability. The absence of mechanisms for uncertainty quantification (UQ) has been widely recognized as a major barrier to the safe, trustworthy, and explainable deployment of ARRD systems in medical imaging [12, 13]. Without reliable uncertainty estimates, clinicians cannot distinguish trustworthy model outputs from cases requiring further review [14].

To address this limitation, a growing body of work has explored uncertainty-aware approaches for ARRD and medical VLMs. Prior studies have investigated uncertainty through a variety of perspectives, including incorporating uncertainty into report generation models, analyzing the stability or consistency of generated text, and applying auxiliary mechanisms to assess the reliability of generated findings [15–17]. These efforts suggest that UQ can be informative for identifying potentially unreliable reports. However, most existing approaches rely on heuristic or model-dependent confidence measures and do not provide statistically grounded guarantees that link uncertainty estimates to error control in deployment settings.

Conformal prediction (CP) offers a model-agnostic, post-hoc, and distribution-free statistical framework for UQ with guaranteed coverage [18, 19]. CP has demonstrated increasing utility in medical imaging tasks such as classification, detection and segmentation, enabling reliable uncertainty estimation without modifying underlying model architectures [20-22]. More recently, CP has begun to be explored in multimodal and vision–language settings, including applications related to factuality assessment in VLMs [17].

In this study, we propose CONRep, a framework designed to enhance the reliability of VLM-based ARRD. The proposed approach aims to provide clinicians with not only the diagnostic impressions derived from imaging data but also explicit indicators of the model's confidence in each finding. By embedding statistically principled UQ into the report generation process and translating it into clear, clinically interpretable language, this work seeks to promote greater transparency, strengthen clinician confidence, and facilitate the safe integration of ARRD systems into routine radiological practice.

**Methods**

We proposed CONRep, a CP framework for UQ in VLM-based chest radiograph interpretation. We applied CONRep in two separate pipelines: (1) a label-based pipeline that quantifies uncertainty for predefined findings as binary outputs, and (2) a sentence-based pipeline that quantifies uncertainty for free-text impression drafting using image–report semantic alignment.

**1. Label-based uncertainty quantification**

The label-based pipeline formulates ARRD as a set of binary classification outputs, where each predefined clinical finding is assessed independently for presence or absence. We evaluated this pipeline on the ChestX-Det10 dataset, which provides radiologist-labeled chest radiographs across ten thoracic findings [23]. For each image–label pair, an alignment estimate is obtained and subsequently processed using CP to derive uncertainty-aware prediction sets. Two model families were evaluated: a decoder-only VLM and a contrastive image–text VLM.

**1.1. Decoder-only vision–language model**

**Model and prompting strategy**

We used MedGemma, a multimodal decoder-only VLM, to perform label-level inference. For each image and each selected pathology label, the model was prompted in a constrained binary question-answering format. The system prompt explicitly instructed the model to behave as an expert radiologist and to respond strictly with either *"Yes"* or *"No"*, thereby avoiding free-form generation and ensuring deterministic label outputs.

**Probability extraction**

Although MedGemma produces discrete outputs, softmax outputs were derived from the model at token-level. Specifically, after generation, the softmax distribution of the final token was computed, and the probabilities associated with the tokens *"Yes"* and *"No"* were extracted. To remove probability mass assigned to irrelevant tokens, the probability of label presence was normalized as:

$$p_{yes} = \frac{P\ (Yes)}{P\ (Yes) + P\ (No)}$$

This yielded a continuous confidence score $p_{yes} \in [0,1]$ for each label and image.

## 1.2. Contrastive vision–language model

**Model and scoring formulation**

We additionally evaluated a contrastive approach using BiomedCLIP, which jointly embeds images and text into a shared semantic space. For each pathology label, two complementary textual prompts were constructed:

- *"a chest X-ray of a person with [condition]"*
- *"a chest X-ray of a person without [condition]"*

Cosine similarity measures how strongly an image aligns with each text prompt, but the absolute value can vary due to factors. Therefore, we compute the difference between similarities to the "with [condition]" and "without [condition]" prompts to obtain a condition-specific score that reflects relative evidence for pathology presence. This reduces sensitivity to global shifts and improves interpretability.

**Probability estimation**

The probability of label presence was obtained by applying a sigmoid transformation to the similarity difference:

$$p_{yes} = \sigma(C_{with} - C_{without})$$

where $C$ denotes the cosine similarity between image and text embeddings. This formulation provides a smooth, interpretable probability estimate without requiring generative decoding.

## 1.3. Ground truth extraction

Ground-truth (GT) labels were obtained from reference annotations associated with each imaging study. When available, structured labels provided by radiologists were used directly. In the absence of structured annotations, labels were automatically extracted from radiology reports using established natural language processing–based label extraction methods [24-26]. All labels were binarized to indicate the presence or absence of each clinical finding.

## 1.4. Conformal prediction for label-based outputs

**Calibration and test split**

For both MedGemma and BiomedCLIP outputs, the dataset was randomly divided into calibration (30%) and test (70%) subsets using a fixed random seed.

## Conformal prediction

We used CP to quantify uncertainty in binary prediction. The objective of CP is to meet the following equation:

$$1 - \alpha \leq P(Y \in C(xi)) \leq 1 - \alpha + \frac{1}{n+1}$$

Where $\alpha$ represents the standard error, $Y$ denotes the label true value, $C(xi)$ indicates the prediction set, and $n$ refers to the size of the calibration set.

## Nonconformity scores

For each label $l$, normalized probabilities were converted into nonconformity scores (NCSs):

- For positive cases:

$$NCS^{(1)} = 1 - p_l$$

- For negative cases:

$$NCS^{(0)} = p_l$$

## Calibration nonconformity ordering and threshold selection

For each label $l$, calibration NCSs were computed on the calibration subset and sorted in ascending order. Given a target $\alpha$ (we evaluated $\alpha \in \{0.05, 0.10, 0.20\}$), the NCS threshold $\tau_l$ was selected as the ($\alpha$)-quantile of the calibration NCS distribution:

$$\tau_l = Q_{1-\alpha}(NCS_l)$$

## Prediction sets and uncertainty definition

On the test set, for each case $x$ and label $l$, NCSs were computed for both potential outcomes $y \in \{0,1\}$, i.e., $NCS_l^{(1)}(x)$ and $NCS_l^{(0)}(x)$. The CP set was then constructed as:

$$C_l(x) = \{y \in \{0,1\} : NCS_l^{(y)}(x) \leq \tau_l\}.$$

Thus, a class $y$ was included in the prediction set if its NCS was less than or equal to the conformal threshold $\tau_l$.

Predictions were categorized as:

- **Certain:** $C_l(x) = \{0\}$ or $C_l(x) = \{1\}$
- **Uncertain:** $C_l(x) = \{0,1\}$ or $C_l(x) = \emptyset$

Coverage (portion of cases in test set contain true prediction in the prediction set), certainty rates, and classification performance metrics (accuracy, F1, area under the receiver operating characteristic curve (AUROC), area under the precision-recall curve (AUPRC), sensitivity, specificity) for certain and uncertain subgroups were computed per each label. Figure 1 shows CONRep workflow as a classifier.

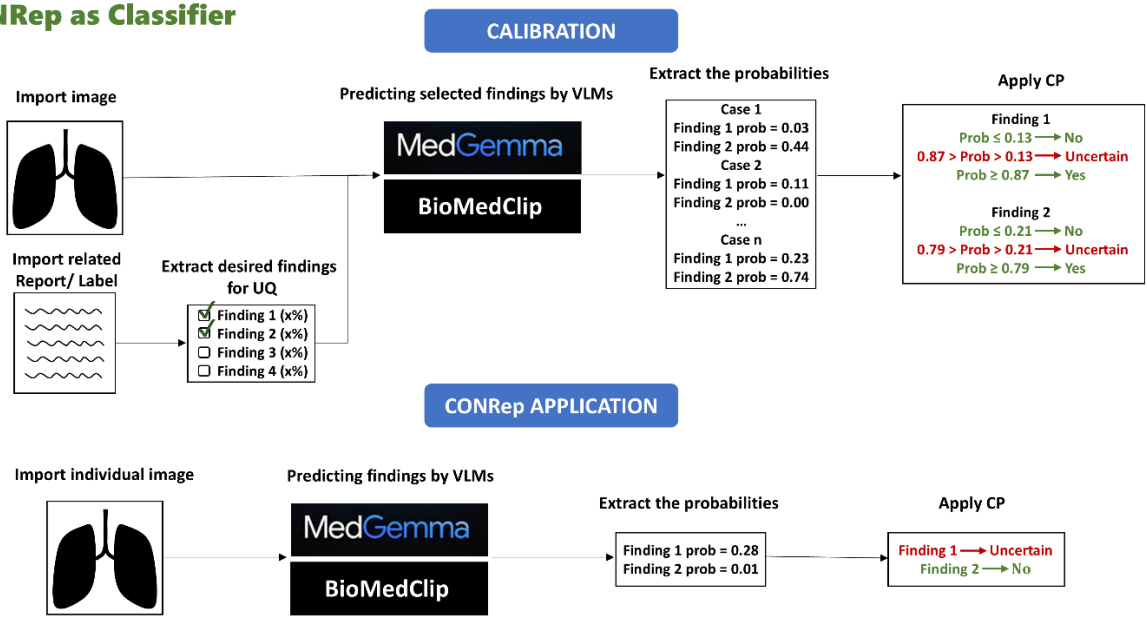

**Figure 1. CONRep as classifier workflow.** Chest X-ray images are processed by two VLM paradigms: a decoder-only VLM (MedGemma) using constrained yes/no prompting and a contrastive VLM (BiomedCLIP) using image–text similarity to estimate label probabilities. A calibration set is used to compute NCS and derive conformal thresholds at predefined significance levels. These thresholds are then applied to the test set to produce prediction sets, enabling classification outputs to be categorized as certain (high confidence) or uncertain (low confidence) with statistical coverage guarantees.

## 2. Sentence-based uncertainty quantification

The sentence-based pipeline evaluates uncertainty at the report level, rather than at the level of individual findings. This approach reflects real-world radiology workflows, where models generate free-text instead of structured labels. We evaluated this pipeline on the Open-I chest X-ray dataset, which includes paired chest radiographs and radiologist-written reports with impression sections used as GT [27].

### 2.1. Impression generation with decoder-only VLM

MedGemma was prompted in a few-shot setting to generate only the impression section of a chest radiograph report. The prompt included multiple curated examples pairing finding and impression text, and the model was instructed to produce concise, clinically appropriate impressions. Deterministic decoding was used to ensure reproducibility.

## 2.2. Image–text semantic similarity with contrastive VLM

For each case, BiomedCLIP embeddings were computed for the:

- frontal chest X-ray image
- generated impression text

The cosine similarity between these embeddings was used as a scalar confidence score reflecting image–text alignment. Similarity values were rescaled to the [0, 1] interval via min–max normalization and treated as pseudo-probability estimates.

## 2.3. Conformal prediction on report-level scores

### Calibration strategy

The dataset was randomly split into calibration and test subsets. During calibration, all samples were treated as belonging to a single positive class representing adequate image–text alignment. The NCS was defined as:

$$NCS(x) = 1 - p_{alignment}(x)$$

A NCS threshold corresponding to $1 - \alpha$ was estimated from the calibration distribution.

### Test-time uncertainty labeling

On the test set, prediction sets were constructed for binary alignment outcomes, and reports were classified as certain, uncertain, and highly uncertain.

- **Certain:** $C_l(x) = \{1\}$
- **Uncertain:** $C_l(x) = \{0,1\}$
- **Highly Uncertain:** $C_l(x) = \{0\}$

## 2.4. Text-Level Evaluation and Statistical Analysis

To assess semantic fidelity, BiomedCLIP text embeddings were computed for both GT impressions and VLM generated reports/impressions (VLMGR), and their cosine similarity was measured. Comparisons were performed between three subgroups (certain, uncertain, and highly uncertain). Also, coverage reported as cases contain true prediction (label 1) in the prediction set. Figure 2 shows CONRep workflow as a report generator.

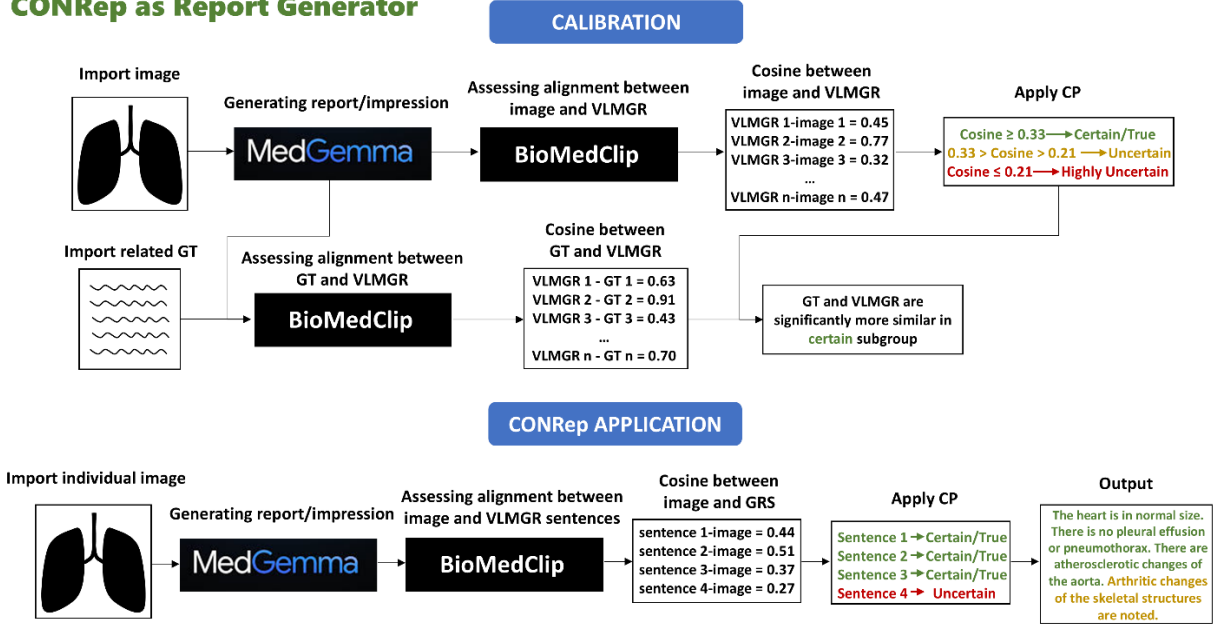

**Figure 2. CONRep as report generator workflow.** Images are input to a VLM for report generation. Image–VLMGR alignment is measured via cosine similarity using a contrastively trained VLM. CP thresholds derived from calibration data are applied to test samples to identify certain, uncertain, and highly uncertain reports.

**Statistical analysis**

Normality was assessed using the Shapiro–Wilk test. Depending on distributional assumptions, either Welch's *t*-test or the Mann–Whitney *U* test was applied. Correlation between variables was evaluated using Pearson and Spearman coefficients.

**Experimental results**

**1. CONRep as classifier (label base)**

Experiments were conducted using the ChestX-Det10 dataset, which contains 3,001 chest radiographic studies independently annotated by three board-certified radiologists. Each study was labeled for the presence or absence of ten thoracic conditions: atelectasis, calcification, consolidation, effusion, emphysema, fibrosis, fracture, mass, nodule, and pneumothorax. Table 1 summarizes the per-label performance of MedGemma and BiomedCLIP on the test set. Across the ten conditions ($n = 10$), the spearman correlation between the decoder-only and contrastive models' AUROC values was 0.936 ($P < 0.01$).

|  | Medgemma predictions | | | | | BioMedClip predictions | | | | |
|---|---|---|---|---|---|---|---|---|---|---|
|  | Acc | AR | AP | Sen | Spe | Acc | AR | AP | Sen | Spe |
| Consolidation | 0.77 | 0.84 | 0.81 | 0.74 | 0.79 | 0.73 | 0.80 | 0.74 | 0.76 | 0.71 |
| Effusion | 0.81 | 0.88 | 0.83 | 0.79 | 0.82 | 0.76 | 0.83 | 0.75 | 0.74 | 0.77 |
| Fibrosis | 0.75 | 0.62 | 0.22 | 0.35 | 0.81 | 0.72 | 0.58 | 0.19 | 0.34 | 0.78 |
| Fracture | 0.82 | 0.68 | 0.23 | 0.33 | 0.88 | 0.35 | 0.59 | 0.13 | 0.86 | 0.28 |
| Nodule | 0.56 | 0.66 | 0.25 | 0.65 | 0.55 | 0.80 | 0.54 | 0.15 | 0.18 | 0.87 |
| Atelectasis | 0.49 | 0.63 | 0.11 | 0.73 | 0.46 | 0.89 | 0.53 | 0.09 | 0.02 | 0.98 |
| Calcification | 0.77 | 0.62 | 0.09 | 0.41 | 0.79 | 0.88 | 0.41 | 0.05 | 0.03 | 0.92 |
| Pneumothorax | 0.83 | 0.86 | 0.25 | 0.70 | 0.83 | 0.76 | 0.74 | 0.18 | 0.53 | 0.78 |
| Emphysema | 0.75 | 0.87 | 0.22 | 0.84 | 0.75 | 0.91 | 0.82 | 0.21 | 0.47 | 0.92 |
| Mass | 0.56 | 0.75 | 0.19 | 0.80 | 0.55 | 0.69 | 0.73 | 0.20 | 0.67 | 0.70 |

**Table 1.** Generative (MedGemma) and contrastively (BioMedClip) VLMs performance on CXR pathologies classification. Acc: accuracy, AR: AUROC, AP: AUPRC, Sen: sensitivity, and Spe: specificity.

Using CONRep, test cases were further stratified into certain and uncertain subsets. Table 2 reports the proportions of certain and uncertain cases and the corresponding performance metrics for each model across three levels of $\alpha$. For both models, the certain subset achieved significantly higher performance for consolidation, effusion, and pneumothorax across all evaluated $\alpha$ values ($P < 0.001$). For MedGemma, this performance difference was also significant for fibrosis, nodule, and emphysema. For BiomedCLIP, the improvement in certain cases reached statistical significance for nodule at $\alpha = 0.05$ and $\alpha = 0.10$.

|  | Label | Certainty | Alpha = 0.05 | | | Alpha = 0.1 | | | Alpha = 0.2 | | |
|---|---|---|---|---|---|---|---|---|---|---|---|
|  |  |  | Ratio | AUC | P val | Ratio | AUC | P val | Ratio | AUC | P val |
| MedGemma Predictions | Consolidation | Certain | 0.46 | 0.90 | <0.01 | 0.61 | 0.89 | <0.01 | 0.80 | 0.87 | <0.01 |
|  |  | Uncertain | 0.54 | 0.74 |  | 0.39 | 0.68 |  | 0.20 | 0.63 |  |
|  | Effusion | Certain | 0.66 | 0.93 | <0.01 | 0.74 | 0.92 | <0.01 | 0.96 | 0.89 | <0.01 |
|  |  | Uncertain | 0.34 | 0.67 |  | 0.26 | 0.64 |  | 0.04 | 0.56 |  |
|  | Fibrosis | Certain | 0.10 | 0.76 | <0.01 | 0.16 | 0.72 | 0.01 | 0.24 | 0.70 | 0.01 |
|  |  | Uncertain | 0.90 | 0.59 |  | 0.84 | 0.58 |  | 0.76 | 0.58 |  |
|  | Fracture | Certain | 0.07 | 0.76 | 0.13 | 0.13 | 0.74 | 0.19 | 0.42 | 0.68 | 0.52 |
|  |  | Uncertain | 0.93 | 0.67 |  | 0.87 | 0.66 |  | 0.58 | 0.65 |  |
|  | Nodule | Certain | 0.20 | 0.77 | <0.01 | 0.41 | 0.73 | <0.01 | 0.79 | 0.68 | <0.01 |
|  |  | Uncertain | 0.80 | 0.58 |  | 0.59 | 0.55 |  | 0.21 | 0.48 |  |
|  | Atelectasis | Certain | 0.45 | 0.65 | 0.08 | 0.56 | 0.66 | 0.01 | 0.79 | 0.64 | 0.06 |
|  |  | Uncertain | 0.55 | 0.59 |  | 0.44 | 0.56 |  | 0.21 | 0.55 |  |
|  | Calcification | Certain | 0.09 | 0.64 | 0.86 | 0.19 | 0.60 | 0.57 | 0.30 | 0.58 | 0.37 |
|  |  | Uncertain | 0.91 | 0.62 |  | 0.81 | 0.64 |  | 0.70 | 0.64 |  |
|  | Pneumothorax | Certain | 0.56 | 0.90 | <0.01 | 0.72 | 0.90 | <0.01 | 0.83 | 0.88 | <0.01 |
| | | Uncertain | 0.44 | 0.71 | | 0.28 | 0.63 | | 0.17 | 0.55 | |
| | Emphysema | Certain | 0.40 | 0.96 | <0.01 | 0.56 | 0.93 | <0.01 | 0.91 | 0.88 | <0.01 |
| | | Uncertain | 0.60 | 0.75 | | 0.44 | 0.74 | | 0.09 | 0.48 | |
| | Mass | Certain | 0.25 | 0.73 | 0.28 | 0.52 | 0.79 | 0.05 | 1.00 | 0.75 | - |
| | | Uncertain | 0.75 | 0.67 | | 0.48 | 0.67 | | 0.00 | - | |
| **BioMedClip Predictions** | Consolidation | Certain | 0.38 | 0.85 | <0.01 | 0.53 | 0.84 | <0.01 | 0.83 | 0.82 | <0.01 |
| | | Uncertain | 0.62 | 0.69 | | 0.47 | 0.66 | | 0.17 | 0.54 | |
| | Effusion | Certain | 0.32 | 0.90 | <0.01 | 0.45 | 0.88 | <0.01 | 0.63 | 0.87 | <0.01 |
| | | Uncertain | 0.68 | 0.74 | | 0.55 | 0.71 | | 0.37 | 0.66 | |
| | Fibrosis | Certain | 0.06 | 0.71 | 0.06 | 0.11 | 0.64 | 0.24 | 0.22 | 0.59 | 0.60 |
| | | Uncertain | 0.94 | 0.56 | | 0.89 | 0.56 | | 0.78 | 0.57 | |
| | Fracture | Certain | 0.38 | 0.57 | 0.90 | 0.86 | 0.58 | 0.39 | 0.94 | 0.59 | 0.89 |
| | | Uncertain | 0.62 | 0.58 | | 0.14 | 0.53 | | 0.06 | 0.60 | |
| | Nodule | Certain | 0.03 | 0.81 | 0.01 | 0.2 | 0.65 | 0.01 | 0.32 | 0.56 | 0.28 |
| | | Uncertain | 0.97 | 0.53 | | 0.8 | 0.53 | | 0.68 | 0.51 | |
| | Atelectasis | Certain | 0.03 | 0.50 | 0.80 | 0.05 | 0.32 | 0.04 | 0.18 | 0.51 | 0.77 |
| | | Uncertain | 0.97 | 0.53 | | 0.95 | 0.52 | | 0.82 | 0.48 | |
| | Calcification | Certain | 0.05 | 0.51 | 0.42 | 0.08 | 0.57 | 0.13 | 0.14 | 0.57 | 0.03 |
| | | Uncertain | 0.95 | 0.40 | | 0.92 | 0.40 | | 0.86 | 0.40 | |
| | Pneumothorax | Certain | 0.38 | 0.81 | <0.01 | 0.49 | 0.82 | <0.01 | 0.66 | 0.80 | <0.01 |
| | | Uncertain | 0.62 | 0.63 | | 0.51 | 0.59 | | 0.34 | 0.53 | |
| | Emphysema | Certain | 0.29 | 0.82 | 0.74 | 0.39 | 0.81 | 0.48 | 0.59 | 0.80 | 0.43 |
| | | Uncertain | 0.71 | 0.78 | | 0.61 | 0.75 | | 0.41 | 0.75 | |
| | Mass | Certain | 0.14 | 0.77 | 0.08 | 0.15 | 0.78 | 0.06 | 0.55 | 0.76 | 0.05 |
| | | Uncertain | 0.86 | 0.66 | | 0.85 | 0.66 | | 0.45 | 0.60 | |

**Table 2. CONrep as Classifier Results**. Generative (MedGemma) and contrastively (BioMedClip) VLMs performance on CXR pathologies classification. Acc: accuracy, AR: AUROC, AP: AUPRC, Sen: sensitivity, and Spe: specificity.

## 2. Sentence based

We applied CONRep to the Open-I chest X-ray dataset, which includes 3660 paired chest radiographs and radiologist reports (GT) containing impression sections. On the test set, the cosine similarity between image-VLMGR showed a significant correlation with the cosine similarity between GT and VLMGR (Figure 3, P<0.001). Moreover, the text-level similarity between GT and VLMGR was significantly higher for certain cases across all three alphas (Table 3, P<0.001).

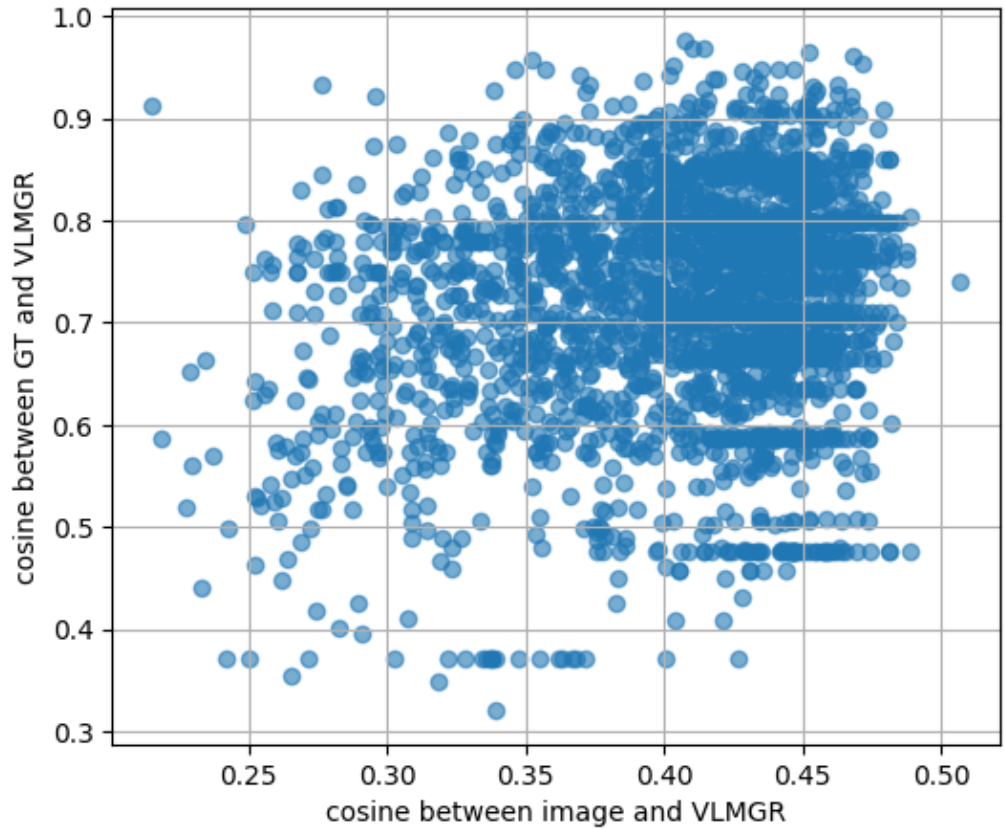

**Figure 3. Correlation between image-VLMGR and GT-VLMGR cosines.** Cosine similarity between image-VLMGR showed a significant correlation with the cosine similarity between GT and VLMGR (P<0.001).

| Alpha | NCS threshold | Coverage | Certain cases | | Uncertain cases | | Highly uncertain cases | | P value |
| --- | --- | --- | --- | --- | --- | --- | --- | --- | --- |
| | | | Ratio | Mean | Ratio | Mean | Ratio | Mean | |
| **0.05** | 0.695 | 0.939 | 52.4% | 0.738 | 41.6% | 0.722 | 6.1% | 0.654 | <0.001 |
| **0.1** | 0.611 | 0.890 | 67.8% | 0.737 | 21.2% | 0.714 | 11% | 0.666 | <0.001 |
| **0.2** | 0.486 | 0.815 | 78.8% | 0.734 | 2.7% | 0.728 | 18.5% | 0.681 | <0.001 |

**Table 3. CONRep as Report Generator Results.** Cosine similarity between GT and VLMGR was significantly higher for certain cases across all three alphas (P < 0.001). Mean: Mean of cosine similarities of GTs and VLMGRs in the test set.

**Discussion**

In this study, we introduce CONRep, a CP–based framework for quantifying uncertainty in VLM radiology report generation. We demonstrate that CONRep can be applied at both the label level and the sentence level to stratify model outputs into high confidence (certain) and low confidence (uncertain) subsets. Across both evaluation settings, outputs classified as certain showed consistently higher agreement with reference annotations and GT impressions, supporting the use of CP to improve the transparency and trustworthiness of ARRD.

In the label-based setting, each VLM (generative or contrastive) produces a probability estimate for the two possible outcomes ($y \in \{0,1\}$) for each finding. These probabilities are converted into NCS, and a label-specific threshold is determinate from the calibration set at a chosen $\alpha$ level. At the test set, the prediction set includes all possible labels whose NCSs fall below this threshold, and flagging cases into certain and uncertain subgroups.

Our results suggest that CP-based stratification is more effective when the baseline model performance for a given finding is adequate. This behavior is illustrated by MedGemma's performance on two representative findings with substantially different difficulty: consolidation and calcification. MedGemma achieved notably higher performance for consolidation (F1 = 0.74; AUROC = 0.84) than for calcification (F1 = 0.15; AUROC = 0.62). In the consolidation task, CP consistently separated test cases into certain and uncertain subsets with significantly higher performance in the certain subset across all evaluated $\alpha$ values (Table 2). In contrast, for calcification, no statistically significant performance differences were observed between the certain and uncertain subsets across the same $\alpha$ levels. This pattern is expected: when predicted probabilities are poorly calibrated or close to random, the induced NCS contain limited discriminative information, and CP tends to produce a large proportion of uncertain cases without meaningfully isolating a high-quality subset. Therefore, CP does not "repair" weak predictors; rather, it provides a principled mechanism to quantify and expose their unreliability.

These findings also highlight that $\alpha$ selection should be label-dependent, rather than fixed globally. Smaller $\alpha$ values impose stricter coverage requirements and generally increase the size of prediction sets, leading to a higher proportion of uncertain cases. For labels with strong baseline performance, a smaller $\alpha$ may be appropriate, as it yields more conservative uncertainty estimates while still preserving a usable proportion of certain predictions. Conversely, for labels with weaker baseline performance, using a larger $\alpha$ may be more clinically practical because overly strict settings can result in an excessively high uncertain-case rate (e.g., fibrosis), limiting the utility of the model in clinical workflows. In this sense, $\alpha$ acts as a controllable operating parameter that trades off coverage guarantees against decisiveness, and optimal selection may vary across findings depending on baseline detectability and clinical risk tolerance.

For the sentence-level CONRep workflow, we quantify uncertainty using an image–report semantic alignment score. Specifically, we compute the cosine similarity between the image embedding and the VLM-generated report embedding (image–VLMGR cosine similarity) using BiomedCLIP. This similarity score is then mapped to [0, 1] via min–max normalization and treated as a pseudo-probability of image–report alignment, where values closer to 1 indicate stronger consistency between the generated text and the image. In this setting, nonconformity is defined as:

$$NCS(x) = 1 - p_{alignment}(x)$$

so that high alignment corresponds to low NCS and low similarity corresponds to high NCS. This formulation is appropriate because the generated report is intended to describe the input image; therefore, misalignment between the generated report and the image should yield high NCS.

In both the label-based and sentence-based settings, we classified cases with $C_l(x) = \{1\}$ as certain, and cases with $C_l(x) = \{0,1\}$ as uncertain. In the label-based setting, we also treated $C_l(x) = \{0\}$ as certain, consistent with standard conformal classification. However, in the sentence-based setting, we interpreted $C(x) = \{0\}$ as highly uncertain, because it indicates that, at confidence level $1 - \alpha$, the model supports non-alignment between the image and generated report. Clinically, this outcome corresponds to a potentially unreliable report that conflicts with the image content and should be prioritized for radiologist review.

Few studies have applied CP to quantify uncertainty in VLM generated medical reports [28, 29]. Prior work has used CP either to calibrate VLM outputs for downstream tasks or to filter low-reliability claims to control factuality risk [28, 29]. In contrast, CONRep applies CP to both finding-level predictions and report-level image–text alignment, producing clinically interpretable certain/uncertain outputs. Across both settings, high-confidence cases achieved significantly better performance than low-confidence cases, confirming that CP effectively identifies more reliable outputs.

The label-based and impression-based implementations of CONRep address complementary aspects of report reliability. Label-based CONRep provides calibrated existence estimates for predefined findings, but it does not capture richer clinical attributes such as location, extent, or associated descriptors. In contrast, impression-based CONRep operates on free-text outputs and can preserve these contextual features (e.g., laterality, distribution, severity), which are often clinically important. However, label-based CONRep produces label-specific NCS thresholds, yielding statistical coverage guarantees for each individual finding, whereas impression-based CONRep relies on a global alignment-based NCS threshold, providing a single coverage guarantee aggregated across all findings rather than selective finding-level control. These approaches are therefore best viewed as complementary: the label-based pipeline can indicate which findings are likely present with calibrated confidence, while the impression-based pipeline can provide a richer narrative explanation of those findings; combining both may yield a practical uncertainty-aware workflow in which CONRep identifies reliable findings and the decoder model contextualizes them in report form.

In the label-based setting, CONRep was applied independently using decoder-only and contrastive VLMs. Although MedGemma showed higher performance for several findings, the goal of this study was not to benchmark architectures, particularly because model release timing and training data differ across paradigms. Instead, we observed strong correlation between the models' per-label performance, indicating that both paradigms tend to be strong on the same findings and weak on the same findings, which may reflect intrinsic detectability differences and dataset-related limitations. For example, some conditions such as effusion are visually salient and consistently documented, whereas others (e.g., fracture) are more subtle and may be underreported or inconsistently represented in GT annotations, which can reduce apparent performance for both models and limit the ability of CP to isolate a clearly higher-confidence subset.

CP offers several practical advantages for UQ in medical imaging report generation. It is model-agnostic, can be applied post hoc without retraining, requires minimal additional computation, and provides statistical coverage guarantees. Also, CP can be integrated across diverse VLM architectures (e.g., decoder-only and contrastive models) because it operates on model outputs rather than requiring structural changes to the underlying model.

Within radiology workflows, CONRep can be integrated as an uncertainty-aware reporting assistant that generates reports accompanied by explicit confidence levels, enabling clinicians to prioritize manual review for low-confidence outputs while using high-confidence outputs as a rapid secondary opinion. Beyond VLM-generated text, the same framework could be applied to radiologist-authored reports to highlight sentences or findings with low image–text alignment, supporting targeted double checking and potentially reducing oversight in high-volume settings.

This study has limitations. Experiments were performed on modest-sized, retrospective public datasets limited to chest radiographs, and generalizability to other modalities and complex examinations remains uncertain. Also, the data reflect single-source distributions and were not externally validated across institutions or acquisition settings, so robustness under domain shift was not assessed. Future work will extend CONRep to additional imaging modalities and multi-sequence studies, and evaluate robustness under multi-center external validation with prospective deployment settings.

**Conclusion**

CONRep provides a statistically valid approach for having uncertainty-aware radiology report drafts using CP. By applying CP to both label and sentence level VLM report generation, CONRep stratifies model outputs into high-confidence and low-confidence subsets with coverage guarantees. Across two VLM paradigms, high-confidence outputs consistently demonstrated significantly higher agreement with reference labels and GT impressions. These results indicate that CP can effectively expose unreliable predictions without modifying underlying model architectures. CONRep therefore represents a lightweight, model-agnostic framework that can support safer integration of VLM-based reporting systems into clinical radiology workflows.

**Declarations**

AI assistant tools

We used ChatGPT-5 to enhance grammar and manuscript content. All the authors reviewed and edited the content to avoid plagiarism.

Acknowledgements

The authors declare that no funds, grants, conflict of interest, or other support were received during the preparation of this manuscript.